\documentclass[twocolumn, showpacs, superscriptaddress, amsmath,amsfonts,prb]{revtex4}
\usepackage{graphicx}
\begin{document}
\title{Nucleation of plasmonic resonance nanoparticles}

\author{Victor G. Karpov}
	\affiliation{Department of Physics and Astronomy, University of Toledo, Toledo, OH 43606, USA}
	\email{victor.karpov@utoledo.edu}
\author{Marco Nardone}
	\affiliation{Department of Environment and Sustainability, Bowling Green State University, Bowling Green, OH 43403, USA}
\author{Nicholas I. Grigorchuk}
	\affiliation{Bogolyubov Institute for Theoretical Physics, NAS of Ukraine - 14-b Metrologichna Str., Kyiv, 03143, Ukraine}

\date{\today}

\begin{abstract}
We predict the electromagnetic field driven nucleation of nanoparticles that provide plasmonic oscillations in resonance with the field frequency. The oscillations assume a phase that maximizes the particle polarization and energy gain due to nucleation. We derive closed form expressions for the  corresponding nucleation barrier and particle shape vs. field frequency and strength, metal plasma frequency, conductivity, and the host dielectric permittivity.  We show that the plasmonic polarization allows for nucleation of particles that would not be stable in zero field.
\end{abstract}
\maketitle

Accelerated nucleation in response to laser or dc electric fields has been observed in a number of systems.\cite{garetz1996,nucrateE,liubin1997,nardone2012}  The phenomenon is typically attributed to the lowering of the nucleation barrier by the field-induced polarization of the new phase particle.  Similar observations have been reported for nanoparticles of various shapes.\cite{lin2011,kim2008,qiu2002,miura2011}  Existing theoretical models \cite{nardone2012,kaschiev2000, warshavsky1999,isard1977} consider an essentially static field.  However, for ac fields, especially near plasmonic resonances,  frequency dependent effects can strongly modify the nucleus polarization and exponentially change the nucleation rate.

This Letter introduces the concept of plasmonic mediated nucleation.  The underlying idea is that the nucleated particle provides a narrow resonance and phase of plasmonic oscillations that maximize its polarizability in the external electromagnetic field; that makes such particles energetically more favorable and lowers their nucleation barriers.  As a result, an ac field will favor the nucleation of a particular geometry, such that the plasmonic frequency of the particle is in resonance with the field frequency.  Furthermore, we will show that the plasmonic related energy gain can be significant enough to change the phase equilibrium, thereby triggering nucleation of particles that would not form in zero field.

Our consideration is based on the classical nucleation theory (CNT) \cite{landau1980,landau2008,kaschiev2000} where the system free energy is approximated by the sum of the bulk and surface contributions of the new phase particle.  Adding an electric field dependent term, $F_E$, the free energy takes the form,
\begin{equation}\label{eq:freeen}
F=F_E+\mu V+\sigma A.
\end{equation}
Here, $\mu$ is the difference in chemical potential (per volume) due to nucleation, and $\sigma$ is the surface tension, $V$ and $A$ are the particle volume and area respectively. The case of $\mu <0$ corresponds to a metastable system in which nucleation is naturally expected without external field; $\mu >0$ describes the case where metal particles are energetically unfavorable in zero field, yet, as shown below, they can appear in a sufficient electromagnetic field.

For a static field, \cite{nardone2012,kaschiev2000, warshavsky1999,isard1977} $F_E$ represents the polarization energy gain similar to that of an induced dipole ${\bf p}=\alpha {\bf E}$ in an electric field ${\bf E}$.  Given a particle of polarizability $\alpha$  in a dielectric material with permittivity $\epsilon$, the polarization induced gain in a static field can be represented as, \cite{kaschiev2000}
\begin{equation}\label{eq:FE0}
F_E=-\epsilon\alpha E^2.
\end{equation}
Although not immediately obvious, the factor $\epsilon$ makes Eq. (\ref{eq:FE0}) different from the energy of a dipole in an external field.  It reflects the contributions from all charges in the system, including those responsible for the field; this factor has been confirmed by several authors. \cite{kaschiev2000,warshavsky1999,isard1977}

In order to introduce some useful notation, we briefly review the case of a spherical metallic nucleus of radius $R$ in a static field.  Using  $\alpha = R^3$, $V=4 \pi R^3/3$, $A=4\pi R^2$, and assuming $\mu <0$, the nucleation barrier $W$ and radius $R$ are determined by the maximum of the free energy in Eq. (\ref{eq:freeen}),
\begin{equation}\label{eq:WR}
W=W_0(1+\xi /2)^{-2}\quad {\rm when}\quad R=R_0(1+\xi /2)^{-1}.
\end{equation}
The corresponding zero-field CNT quantities and the dimensionless field strength parameter $\xi$ are,
\begin{equation}\label{eq:CNT}
W_0=\frac{16\pi}{3}\frac{\sigma ^3}{\mu ^2},\quad R_0=\frac{2\sigma}{|\mu |},\quad \xi=\frac{\epsilon E^2R_0^3}{W_0}.
\end{equation}
Their ballpark values are $W_0\sim 1$ eV, $R_0\sim 1$ nm, and $\xi\ll 1$, say $\xi\sim 10^{-5}$ for a moderate field of $E=30$ kV/cm and $\epsilon =1$.

Since we are considering electromagnetic fields of frequency $\omega\gg \omega _{at}$, where $\omega _{at}\sim 10^{13}$ s$^{-1}$ is the characteristic frequency of atomic vibrations, the induced polarization will be predominantly of electronic origin.  The corresponding part of the electronic energy, which is proportional to $-{\bf p\cdot E}$, will oscillate in time with frequency $\omega$.  According to the standard procedure of adiabatic (Born-Oppenheimer) approximation in solids, the time average of that oscillating energy can be treated as a contribution to the potential energy of the atomic subsystem.  Taking into account the known recipe for time averages, \cite{landau1984} the latter contribution to the free energy of an atomic subsystem can be presented in the form,
\begin{equation}\label{eq:FE}
F_E=-\epsilon\frac{E}{2}\Re(\alpha ),
\end{equation}
where $E$ is now understood as the field amplitude and $\Re (\alpha)$ represents the real part of the polarizability.

\begin{figure}[htb]
\includegraphics[width=0.35\textwidth]{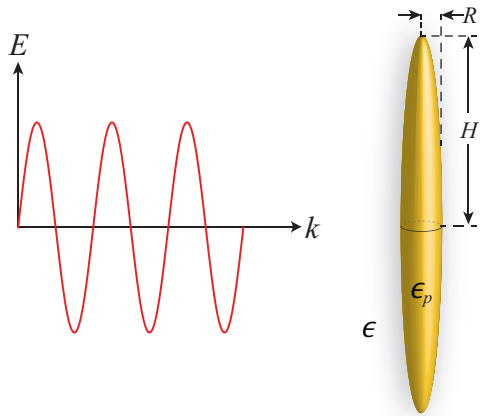}
\caption{Illustration of a plasmonic resonance nanoparticle.  A metallic prolate spheroid of semi-major axis $H$ (aligned with the electric field $E$) and semi-minor axis $R$ is formed with its aspect ratio tuned to the frequency, $\omega$, of the ac field such that $H/R\approx \omega_p/\omega$, where $\omega_p$ is the plasma frequency.  $\epsilon$ and $\epsilon_p$ are the permittivity of the dielectric medium and particle, respectively.\label{Fig:prolate}}
\end{figure}

Prior to the rigorous derivation, we show next that all the major results of this work can be obtained qualitatively [Eqs. (\ref{eq:qualalpha}-\ref{eq:qualbar})].  We start with noting the high polarizability,
\begin{equation}\label{eq:qualalpha}
\alpha\sim (H/R)^2V\gg V,
\end{equation}
of needle-shaped particles, e.g., a prolate spheroid or a cylinder of height $2H$ and radius $R\ll H$ aligned to the field (Fig. \ref{Fig:prolate}). Indeed, the field-induced charges, $\pm q$, induced at the opposite poles are estimated from the balance of forces, $q^2/H^2=qE$, which gives the dipole moment $p\sim Hq\sim EH^3\sim V(H/R)^2E\equiv \alpha E$.

The plasmonic resonance in a needle-shaped metal particle can be qualitatively explained by considering small parallel displacements $x\ll H$ of the free electrons along the cylinder axis. This will result in small charges $q\sim \pm \pi R^2xNe$ on the two poles of the cylinder where $N$ and $e$ are the electron concentration and charge, respectively. The corresponding forces $-qe/H^2$ on individual electrons can be interpreted as the restoring forces $-m\omega ^2x$, which yields the resonant frequency,
\begin{equation}\label{eq:qualres}
\omega _r\sim \omega _p(R/H)\ll \omega _p,
\end{equation}
where $m$ is the electron mass and we have used the standard definition of the plasma frequency,
\begin{equation}\label{eq:omegap}
\omega _p=\sqrt{4\pi Ne^2/m}.
\end{equation}
$\omega _r$ is known as the plasmonic resonance frequency which describes collective oscillations of quasi-free electrons; it has been experimentally observed in light scattering by nanoparticles.\cite{maier2007}

When $\omega _r$ is in resonance with $\omega$, the plasmonic vibrational energy is accumulated in the particle to the extent allowed by its quality factor ($Q$-factor), estimated as $Q\sim \omega\tau \gg 1$ with $\tau$ being the electron relaxation time. Because that energy is proportional to the square of the internal electric field, $E_{int}^2$, one can say that the latter increases the polarizability effectively by the factor $\omega _r\tau$.

Combining the latter with Eq. (\ref{eq:qualalpha}) and (\ref{eq:qualres}) gives the resonant ($\omega\approx\omega _r$) plasmonic enhanced polarizability,
\begin{equation}\label{eq:qualpol}
\alpha _r\sim V(\omega _p/\omega )^2\omega\tau\gg V.
\end{equation}
This gigantic increase in polarizability takes us to the major prediction of this work: an ac field of frequency $\omega$ can drive the nucleation of needle-shaped particles with resonant aspect ratio $H/R \sim \omega _p/\omega\gg 1$.

We note that the first amplification factor  $(\omega _p/\omega )^2\sim (H/R)^2\gg 1$ reflects the needle-shaped geometry of the particle and remains as such in a static field, \cite{nardone2012} while the $Q$-factor enhancement $\omega\tau\gg 1$ is specific to ac fields.  Taking into account the amplification ratio in Eq. (\ref{eq:qualpol}), an approximate result for the ac resonant nucleation barrier can be guessed from the known static result, \cite{nardone2012} or even from Eq. (\ref{eq:WR}), with $\xi\rightarrow  (\omega _p/\omega )^2(\omega\tau )\xi\gg 1$, which yields,
\begin{equation}\label{eq:qualbar}
 W\sim (\omega /\omega _p^3\tau ^2)(W_0/E^2R_0^3)^2W_0.
 \end{equation}

Our rigorous analysis begins with the polarizability of a spheroid:\cite{bohren1983}
\begin{equation}\label{eq:alpha}
\alpha = \frac{V}{4\pi}\frac{\epsilon _p-\epsilon}{\epsilon+n(\epsilon _p- \epsilon)}.
\end{equation}
Here, $\epsilon _p$ is the dielectric permittivity of a metal particle and $\epsilon$ is that of the medium, $n$ is the depolarizing factor, which we take as that of a strongly anisotropic prolate spheroid,
\begin{equation}n\approx (R/H)^2[\ln(2H/R)-1]\equiv (R/H)^2\Lambda\ll 1.\label{eq:ncyl}\end{equation}
Also, we use the dielectric permittivity of a metal,
\begin{equation}\label{eq:metal}\epsilon _p=1-\frac{\omega _p^2}{\omega ^2}+i\frac{\omega _p^2}{\omega ^3\tau}.\end{equation}
$\epsilon$ is assumed to be a real number, and $\tau ^{-1}\ll\omega\ll\omega _p$.

\begin{figure}[htb]
\includegraphics[width=0.45\textwidth]{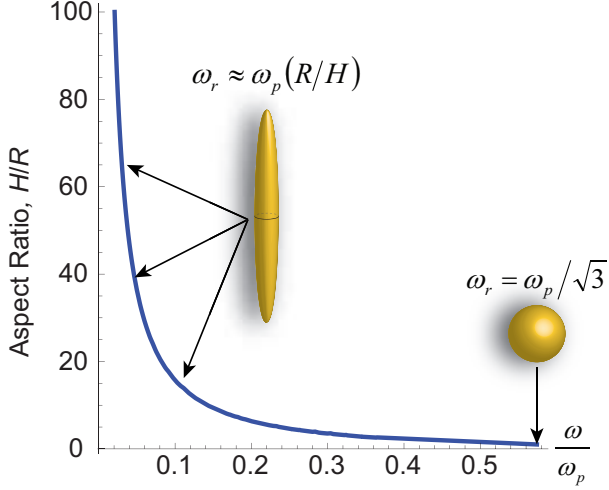}
\caption{Aspect ratio of the metallic nucleus as a function of frequency.  The resonance frequency for a sphere is $\omega_r=\omega_p/\sqrt{3}$, while for a prolate spheroid the resonance frequency depends on the aspect ratio, as shown.\label{Fig:aspectratio}}
\end{figure}

Using Eqs. (\ref{eq:alpha} - \ref{eq:metal}) with Eq. (\ref{eq:FE}) yields,
\begin{equation}\label{eq:realpha}
\Re (\alpha )=\frac{V}{4\pi}\frac{(n-n_{\omega})+bn}{(n-n_{\omega})^2+bn^2},
\end{equation}
where $b=1/(\omega\tau )^2$ and,
\begin{equation}\label{eq:param}
n_{\omega}=\frac{\epsilon\omega ^2}{\omega _p^2+(\epsilon -1)\omega ^2}\approx\frac{\epsilon\omega ^2}{\omega _p^2}\ll 1.
\end{equation}
The polarizability, $\Re (\alpha )$, has a sharp minimum when,
\begin{equation}\label{eq:nres}
n \approx n_{\omega}-\sqrt{b}n_{\omega},
\end{equation}
which reflects the presence of the plasmonic resonance.  Since $n\sim (R/H)^2$ from Eq. (\ref{eq:ncyl}), the expression for $n_{\omega}$ is fully consistent with the qualitative result in Eq. (\ref{eq:qualres}).  The relationship between nanoparticle aspect ratio and the field frequency is illustrated in Fig. \ref{Fig:aspectratio}.

Given the sharpness of the resonance, all other $n$-dependent quantities, in particular the prolate spheroid volume and area,
\begin{equation}\label{eq:volume}
V=\frac{4\pi H^3n}{3\Lambda},\quad A=\pi ^2H^2\sqrt{\frac{n}{\Lambda}}.
\end{equation}
can be evaluated at $n=n_{\omega}$.  That yields,
\begin{equation}\label{eq:LRmin}
\left[\Re (\alpha )\right]_{min}\approx -\frac{H^3}{6}\frac{1}{\Lambda \sqrt{b}},
\end{equation}
fully consistent with our earlier estimate in Eq. (\ref{eq:qualpol}).

\begin{figure}[htb]
\includegraphics[width=0.45\textwidth]{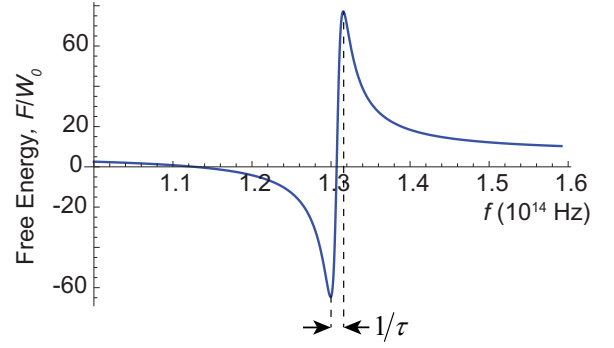}
\caption{Normalized free energy of a prolate spheroidal metallic particle vs. field frequency $f=\omega /2\pi$, with $E=3\times 10^5$ V/cm. The sharp resonance of width $1/\tau$ determines the aspect ratio $H/R\approx \omega_p/\omega\approx 10$.  Parameters values are typical for nucleation in solids: $R_0=1$ nm and $W_0=1$ eV, with $\omega_p=10^{16}$ rad/s, $\tau=10^{-13}$ s, and $\epsilon=1$. \label{Fig:freeprolate}}
\end{figure}

Normalizing the free energy in Eq. (\ref{eq:freeen}) with respect to the classical barrier, it takes the form,
\begin{align}\label{eq:freenorm}
\frac{F}{W_0}&=\frac{H^3}{R_0^3}\left(\frac{\omega }{\omega _p}\right)^2\frac{2\epsilon}{\Lambda}\left[-\frac{E^2 \epsilon R_0^3}{24W_0}\left(\frac{\omega _p}{\omega}\right)^2(\omega\tau )\pm 1\right]\nonumber \\&+\frac{3\pi}{4}\sqrt{\frac{\epsilon}{\Lambda}}\frac{\omega}{\omega _p}\frac{H^2}{R_0^2},
\end{align}
The minus sign in the second term corresponds to the case when the original phase is metastable and, thus, nucleation can occur in zero field.  The plus sign describes the case of a stable original phase, i.e., nucleation is impossible in zero field.  The effect of the resonance on the free energy is illustrated in Fig. \ref{Fig:freeprolate}.

It follows from Eq. (\ref{eq:freenorm}) that the plasmonic resonance mechanism can significantly accelerate nucleation in metastable systems. For the case of stable systems, the field can induce nucleation when,
\begin{equation}\label{eq:fieldcrit}
E> E_c\frac{1}{\sqrt{\omega\tau}}\frac{\omega}{\omega _p},\quad E_c\equiv 2\sqrt{\frac{6 W_0}{\epsilon R_0^3}}.
\end{equation}
Assuming the above ballpark parameter values yields the characteristic field $E_c\sim 10^8$ V/cm.  However, the other multipliers can make the right hand side in the inequality of Eq. (\ref{eq:fieldcrit}) much lower, down to say 10-100 kV/cm, corresponding to a moderate laser power density of $P\sim 10-100$ mW/$\mu$m$^2$.

From this point on we concentrate on the case of field dominated nucleation, which satisfies the condition of Eq. (\ref{eq:fieldcrit}) and implies neglecting the second term in the square brackets of Eq. (\ref{eq:freenorm}).  In minimizing $F$ with respect to $H$ we keep
\begin{equation}\label{eq:Lambda}
\Lambda\approx \ln\left(\frac{2\omega _p}{\sqrt{\epsilon}\omega }\right)-1
\end{equation}
constant. The resulting nucleus semi-major axis and nucleation barrier are, respectively,
\begin{equation}\label{eq:H0}H=\frac{\pi }{4}\sqrt{\frac{\Lambda}{\epsilon}}\left(\frac{E_c}{E}\right)^2\frac{1}{\omega _p\tau}R_0,\end{equation}
\begin{equation}\label{eq:nucbar}W=\frac{\pi ^3}{64}\sqrt{\frac{\Lambda}{\epsilon}}\frac{\omega}{\omega _p^3\tau ^2}\left(\frac{E_c}{E}\right)^4W_0.\end{equation}

The above theory applies most easily to a system close to some phase transition at temperature $T_c$.  That enables one to estimate the chemical potential as $\mu =\mu _0(1-T/T_c)$. Correspondingly, the classical nucleation radius and barrier become,
\begin{equation}\label{eq:param1}
R_0=R_{00}(1-T/T_c)^{-1},\quad W_0=W_{00}(1-T/T_c)^{-2},
\end{equation}
and $E_c=E_{c0}(1-T/T_c)^{1/2}$, where $R_{00}$, $W_{00}$, and $E_{c0}$ are obtained from their definitions in Eqs. (\ref{eq:CNT}) and (\ref{eq:fieldcrit}) with $\mu =\mu _0$; numerically, they are of atomic scale.  The convenience of this model is that it allows macroscopically large $R_0$, consistent with CNT; also, it corresponds to lower a critical field, $E_c$, making the plasmonic nucleation effect easier to observe.

Using the above scaling, the results for particle nucleation length and  barrier turn out to be temperature independent: they retain their form of Eqs. ({\ref{eq:H0}) and (\ref{eq:nucbar}) with the trivial substitutions, $$R_0\rightarrow R_{00},\quad W_0\rightarrow W_{00}, \quad E_c\rightarrow E_{c0}.$$  This is in striking difference with CNT, which predicts a diverging nucleation  radius $R_{0}$ and  barrier $W_0$ towards the phase transition temperature in Eq. (\ref{eq:param1}).

Numerically, there is a broad window wherein the above predicted plasmonic resonance nucleation takes place.  Indeed, the quality factor at the plasma frequency can be represented\cite{ashcroft} as $\omega_p\tau = 160/(\sqrt{Na_B^3}\rho )\sim 10^3$, where $a_B$ is the Bohr radius.  The factor $1/\sqrt{Na_B^3}$ is typically in the range of 5-10, and the resistivity, $\rho$, (in units of $\mu\Omega$ cm) is typically smaller than unity.  However, this estimate may become less optimistic if $\tau$ is additionally reduced by particle surface scattering when the particle dimension is very small. \cite{grigorchuk2012}

We now briefly discuss possible implications of our theory. A system near phase transition under a moderate intensity laser beam would be most suitable for experimental verification.  The predicted high aspect ratio particles are optically accessible and can be identified via the unique features of prolate spheroids in light scattering and absorption. \cite{maier2007,grigorchuk2012} Another unique feature is that the laser beams that create the nanoparticles can simultaneously play the role of optical tweezers.  That can be exploited to observe and control the positions of nucleated metal particles that would be unstable in zero field.

There have been several observations that are possibly consistent with the above predicted mechanism, a few of note include: the formation of metallic nanorods under laser beams, which did not nucleate in zero field; \cite{lin2011} many orders of magnitude acceleration of metal nano-particle formation under laser irradiation; \cite{kim2008} laser induced nucleation of silver nanoparticles in glasses; \cite{qiu2002} and a versatile, comprehensive study of laser induced nano-wire nucleation. \cite{miura2011} In all those cases, different mechanisms may have been responsible for the observations (except, perhaps, for the case\cite{lin2011} that specifically identifies the frequency dependent effect and the possible role of plasmonic excitations).  Further experimental verification is required to attribute the observations to the phenomenon predicted here.  Verifiable features include frequency dependent nanoparticle nucleation and shape, nucleation rates that are exponentially accelerated with laser power, and Arrhenius temperature dependence of the nucleation rate.  It should be noted that, because the laser beam action continues for a certain time, the nucleated particles can evolve further, undergoing growth, which can affect their expected shapes.

Given the abundant current literature on the topic, we can safely omit the discussion of important practical applications of nanoparticle plasmonic resonances.  However, we emphasize that because of its resonant nature, the mechanism introduced in this work can be useful in better controlling nanoparticle sizes, shapes, and distributions.

In conclusion, we have predicted a phenomenon of plasmonic driven nucleation of metallic nanoparticles.  There are many important practical implications.  More work is called upon to describe the growth stage of such resonant particles and relate them to the experimental observations.

\end{document}